\documentclass[conference]{IEEEtran}

\usepackage{amsmath,amssymb,amsfonts}
\usepackage{algorithmic}
\usepackage{graphicx}
\usepackage{enumerate,setspace,xcolor}
\usepackage[boxruled,vlined]{algorithm2e}
\usepackage{xcolor}
\usepackage{multirow}
\usepackage{balance,cite}

\usepackage{float}

\begin{document}

\title{A Distributed Algorithm for Overlapped Community Detection in Large-Scale Networks}

\author{\IEEEauthorblockN{ Dibakar Saha}
\IEEEauthorblockA{\textit{Department of Mathematics} \\
\textit{Indian Institute of Technology Guwahati}\\
dibakar.saha10@gmail.com}
\and
\IEEEauthorblockN{ Partha Sarathi Mandal}
\IEEEauthorblockA{\textit{Department of Mathematics} \\
\textit{Indian Institute of Technology Guwahati}\\
psm@iitg.ac.in}
}

\maketitle             

\begin{abstract}

Overlapped community detection in social networks has become an important research area with the increasing popularity and complexity of the networks. Most of the existing solutions are either centralized or parallel algorithms, which are computationally intensive - require complete knowledge of the entire networks. But it isn't easy to collect entire network data because the size of the actual networks may be prohibitively large. This may be a result of either privacy concerns (users of a social network may be unwilling to reveal their social links) or technological impediments (implementation of an efficient web crawler). Performing in-network computation solves both problems utilizing the computational capability of the individual nodes of the network.  
Simultaneously, nodes communicate and share data with their neighbours via message passing, which may go a long way toward mitigating individual nodes' privacy concerns in the network. 

All the aforementioned concerns motivated us to design a decentralized or distributed technique to detect overlapped communities in a large-scale network. It is desirable because this technique does not offer a single point of failure, and the system as a whole can continue to function even when many of the nodes fail.
To overcome the disadvantages of the existing solutions,
in this paper, we address the overlapped community detection problem for large-scale networks. 
We present an efficient distributed algorithm, named \texttt{DOCD}, to identify the overlapped communities in the network. The efficiency of \texttt{DOCD} algorithm is verified with extensive simulation study on both synthetic and real networks data such as, \texttt{Dolphin}, \texttt{Zachary karate club}, \texttt{Football club}, and \texttt{Facebook ego} networks. We show that \texttt{DOCD} algorithm is capable of keeping the asymptotically same results with the existing classical centralized algorithms \cite{Newman-2004,Blondel-2008,Ghoshal-2017,Jokar-2019,Meena-2015} in terms of community modularity and the number of identified communities. The \texttt{DOCD} algorithm can also efficiently identify the overlapped nodes and overlapped communities with a small number of rounds of communication and computation. 
\end{abstract}

\section{Introduction}
Large-scale networks, such as social networks (e.g., Facebook, Twitter, Linkedin, ResearchGate, Instagram, etc.), consist of a large number of connected individuals or users; typically, the number of nodes is millions, and the number of links between the users is billions in such a social network. The size of such networks is growing with an increasing number of nodes at an enormous rate day by day. Thus, in such networks, the demand for analysis and characterization is increasing rapidly. This analysis provides an immense knowledge for understanding the features of the network entities and other related characteristics.
For instance, the connectivity between users in the social network represents their friendship relation or neighbourhood relation or both. They can belong to a group, but analysis can establish that those 
group of people is from the same school or college or maybe from the same religion, which in turn refers to a {\it community}.
To identify such {\it communities} is one of the techniques by which we can analyze and characterize the networks. 
Mathematically, a network is considered a graph where nodes represent vertices, and links between nodes represent edges of the graph. 
\textit{Community detection} in networks is a process of partitioning the underlying network graph into subgraphs (i.e., communities), which are internally densely and externally sparsely connected. This is a fundamental analysis technique that confers the modular composition of a network.

Community detection has been used in a broad range of applications, such as to find research communities in DBLP databases, to identify functional groups of a particular virus strains for vaccine development, and to classify content on social media sites.
Prior knowledge of communities helps to understand the processes like rumor spreading or epidemic spreading in a network. For example, there is a very high chance of spreading some infectious diseases very fast within a community where community members are physically staying together or very near to each other (e.g., housing complex, town, city, village, etc.) or sharing common places (e.g., market, shopping mall, school, college, rail station, etc.). Therefore, appropriate preventive measures can be applied before spreading, or essential help can be provided to the affected communities.
Similarly, communities allow us to create a large-scale map of a network, making it easier to study and analyze. 

Moreover, in social networks, a person belongs to more than one social groups such as family, friends, colleagues, where each group can be treated as an individual community. Thus, a person can simultaneously associate with as many communities as he wishes. Therefore, if a node belongs to more than one community in the network, it is an \textit{overlapped node}. When those overlapped nodes formed a community, which is termed as \textit{overlapped community}. 
Such overlapped communities are frequently visible in social networks like Facebook, Twitter, etc. Finding overlapped communities refers to the overlapped community detection problem in networks.\\
Designing practical algorithms for overlapped community detection in network graphs is an important and challenging problem. However, centralized or parallel algorithms are computationally intensive and require complete knowledge of the networks, which is not feasible for large-scale networks.
Therefore, an efficient, distributed algorithm to find overlapped communities in large-scale networks is needed. This paper proposes an efficient, distributed algorithm to identify overlapped communities in large-scale networks using local information and message passing.\\
\noindent \textit{Our Contributions:} 
This paper studies distributed overlapped community detection problem in large-scale networks
and makes the following contributions.

\begin{enumerate}
\item We propose a distributed algorithm (\texttt{DOCD}) that can efficiently identify the overlapped communities in large-scale networks. To the best of our knowledge, this is the first distributed algorithm to identify an overlapped community.
\item  The number of nodes in the networks is not an input of the algorithm. It is scalable and robust with respect to the number of nodes in the networks. 
\item  The time and message complexities of the algorithm are $O(n^2m)$ and $O(\mathcal{D})$, respectively, where $m$, $n$, and $\mathcal{D}$ are the number of nodes, edges, and diameter of the underlying network graph. 
 \item We report on the performance of our algorithm through simulation. The efficiency of the \texttt{DOCD} algorithm is verified with extensive simulation study on both synthetic and real networks data such as, \texttt{Dolphin}, \texttt{Zachary karate club}, \texttt{Football club}, and \texttt{Facebook ego} networks. 
\item We show that the \texttt{DOCD} algorithm (with local information) is capable of keeping the asymptotically same results with the existing classical centralized algorithms \cite{Newman-2004,Blondel-2008,Ghoshal-2017,Jokar-2019,Meena-2015} (which need the information of entire networks) in terms of community modularity, number of identified communities, overlapped nodes, and communities. 
\end{enumerate}

The rest of the paper is organized as follows. Section \ref{sec:rw} presents brief literature review.
In Section \ref{sec:problem}, we present the preliminaries and formulate the problem.
Section \ref{sec:algo} presents the overlapped community detection algorithm.
The detailed analysis of the message and time complexity is presented in Section~\ref{sec:complexity}. 
Section \ref{sec:simulation}, evaluates the performance of the proposed method and finally, Section \ref{sec:conclusion}, concludes the paper.

\section{Related Works}
\label{sec:rw}
Several community detection techniques are reported in the literature, which is mostly centralized and parallel solutions.
The community modularity proposed by Girvan \& Newman \cite{Newman-2004} is the widely used \cite{Blondel-2008,Clauset-2004,Lu-2015} metric to measure connectivity among community member, i.e., the structure of a community in an underlying network graph.  
Community merging between any two communities leads to the maximum modularity gain, and this technique has been used in a \texttt{CNM} algorithm \cite{Clauset-2004}. The \texttt{CNM} algorithm is a hierarchical agglomeration algorithm for detecting the community structure. Further, the \texttt{CNM} algorithm has been used in \cite{Blondel-2008,Wakita-2007}.
Several other measurement techniques are proposed   \cite{Prat-2014,Ghoshal-2017} to detect the quality of the community structure. 
For instance, the paper \cite{Prat-2014} presents an algorithm that detects disjoint communities in a network using another community modularity metric called \texttt{WCC}, which uses {\it triangular} structures in the community. 
Ghoshal \cite{Ghoshal-2017} defines another {\it modularity} metric to show the improvement of their algorithm compare to the technique proposed in \cite{Newman-2004,Clauset-2004}.
The authors in \cite{Bai-2017} proposed an iterative search algorithm for community detection that uses
a community description model evaluates the quality of a partition, where the partition is done based on external-link separation among the communities and internal-link compactness within communities. 

Brandes et al. \cite{Brandes-2008} proved that identifying communities with maximum modularity is NP-hard, even for the restricted
version with a bound of two on the number of clusters, i.e., communities, and established a lower bound on the approximation factor.  
Therefore, heuristics are used in practice to allow the processing of large inputs. However, even such heuristics could take a very long time or run out of memory on modern days' computer. Hence, parallel or distributed solutions are very much essential to reduce computation time. An MPI based parallel heuristics for community detection has been proposed in \cite{Lu-2015}. They developed a parallel version of the \texttt{Louvain} method \cite{Blondel-2008} to reduce the time complexity. However, their proposed parallel method requires repetitive tasks such as graph coloring, meta node creation in each iteration. 
The authors in \cite{Staudt-2016} proposed parallel version of the label propagation method \cite{Raghavan-2007}, \texttt{Louvain} method \cite{Blondel-2008}, called as \texttt{PLM}. Next, they extended this \texttt{PLM} method to \texttt{PLMR} method. 
Finally, the authors combined both the \texttt{PLM} and \texttt{PLMR} algorithms to present a two-phase approach called as \texttt{EPP}.
A divisive spectral method has been presented in \cite{Cheng-2016}, where the authors first used a sparsification operation followed by a repeated bisection spectral algorithm to find the community structures.

However, all the above works are based on finding the disjoint communities. Another important direction of the identification of communities is whether the communities are overlapped community or not. This problem has been addressed in \cite{Bandyopadhyay-2015}, and the authors proposed a centralized algorithm that finds the overlapped communities. Furthermore, Said et al. \cite{Said-2018} used a genetic algorithm to find the overlapped communities in social and complex networks. 
The authors in \cite{Meena-2015}, proposed a two-step genetic algorithm to find the overlapped communities. They first encountered the disjoint communities, and from the disjoint communities, the overlapped communities have been identified using the community modularity as the optimization function.
Reihanian et al. \cite{Reihanian-2018} proposed a generic framework to find the overlapping communities in social networks, where paper focused on rating-based social networks. The members within a community have the same topics of interest. The strengths of the relationships between the members are based on the rate of their viewpoints’ unity, where the strengths of connections of intra-communities are much more than those of inter-communities.
All the solutions have been proposed so far for addressing the community detection problem in large-scale networks. However, many of them are centralized or parallel with expensive procedures either requires complete knowledge of the networks or computationally intensive. Using local information, designing practical distributed algorithms for community detection is an important and challenging problem in large-scale networks.

\section{Preliminaries}
\label{sec:problem}

\subsection{Basic Definitions}

\noindent {\bf Network graph:}
A network graph is denoted as $G(V,E)$, which shows interconnections between a set of entities
$V=\{v_1,v_2,\ldots,v_n\}$. Each entity $v_i \in V$ is represented by a node or vertex. Connections between nodes are represented through links or edges set $E$.
Let $n=|V|$ and $m=|E|$ be the number of vertices and number of edges in $G$, respectively.
$G(V, E)$ is a simple undirected graph.

\noindent {\bf Community in a Graph $G(V,E)$:}
We denote a set $C=\{C_1,C_2,\ldots,C_k\}$ which consists of $k$ number of communities, where each community $C_i\in C$, $1\leq i\leq k$, consists of set of nodes of $V$, i.e., $C_i\subseteq V$.
The size of a community $C_i$ is presented by $|C_i|=\lambda$, where $C_i=\{v_1,v_2,\ldots, v_{\lambda}\}$ and $V=C_1\cup C_2\cup\ldots \cup C_k$.

A node may belong to one or more communities.
If a node $v$ belongs to at least two communities such as $C_i$ and $C_j$, i.e., 
$v \in C_i \cap C_j$ then $v$ is treated as an overlapped node.

\noindent {\bf Overlapped node:} A node is said to be an overlapped node if it belongs to at least two communities.

\noindent {\bf Overlapped community:} A community is said to be an overlapped community if it consists of at least one overlapped node.

\noindent {\bf Cluster coefficient:} The \textsl{cluster coefficient} \cite{Said-2018} of a node $v$ is denoted as: 
$CC_v=\frac{2\times \bar{\mu}}{\delta(\delta-1)}$, where $\bar{\mu}$ is the total number of links between neighbours and $\delta$ is the total number of neighbours of $v$.

\noindent {\bf Node Modularity (NM):} The {\it node modularity} of a node $v$ is defined as: 
\begin{equation}
\label{eqn:nm}
NM_v=\frac{(2\times \mu)}{\delta(\delta-1)},
\end{equation}
where $\mu$ is the `total number of links between neighbours within its own community of $v$' and $\delta$ is the total number of neighbours of $v$.

\noindent {\bf Overlapped Node Modularity (ONM):}
The {\it Overlapped Node Modularity} is defined as:
\begin{equation}
\label{eqn:onm}
ONM_v=\frac{(2\times \mu')}{\delta(\delta-1)},
\end{equation}
 where $\mu'$ is the `total number of links between neighbours belong to communities of overlapped node $v$'.
 
For example, let  $v$ be an overlapped node belongs to two different communities $C_i$ and $C_j$. Hence, all the links of its neighbours that belong to $C_i$ and $C_j$ are only be considered for computing $ONM_v$ of $v$. 
The value of $NM_v$ and $ONM_v$ is the same when $v$ is not an overlapped node.

\noindent {\bf Community Modularity:}
Community modularity measures community structures. It quantifies how a node within a community is strongly connected with other nodes in the network. We use this metric to measure the connectivity between nodes and identifying communities.
The \textit{community modularity} of a community $C_i\in C$ is defined as:
 $ CM_{C_i}=\frac{\sum_{j=1}^{\lambda} NM_j}{\lambda}.$
 
\noindent {\bf Overall Community Modularity:} The \textit{overall community modularity} of the set of communities $C$ is defined as:
 $ CM_C=\frac{\sum_{i=1}^{k} CM_{C_i}}{k}.$

\subsection{Distributed Computing Model}
We consider a synchronized communication network which consists of $n$ nodes. Each node has a unique id of an $O(\log~ n)$ bits. Initially, each node knows its own id
and the ids of its neighbours in the network. 
We use \textit{CONGEST} model \cite{Peleg-2000,Molla-2019} of distributed computing, wherein each communication round, every node may
send an $O(\log~n)$-bit message to each of its neighbours.
In this synchronous system, the computation of every node proceeds in rounds. In each round, every node of the network sends messages, and its neighbours receive these messages in the same round. 

We consider the distributed network model as a network graph $G = (V, E)$, where nodes represent processors (or computing entities), and the edges represent communication links among the processes. Nodes communicate through the edges in synchronous rounds.
At the beginning, each node knows following additional information:
\begin{itemize}
\item  $\Gamma(v)$: list of neighbours of a node $v$ and adjacent edges of each neighbour;
\item  $\gamma(v)$:  a list of communities where neighbours of $v$  belongs to. Each entry in this list is represented by a $3$-tuple: $\{c\_id,c\_size,CM\}$, where $c\_id$ is the community ID, $c\_size$ is the number of node in the community and $CM$ is the \textit{community modularity};
\item  $join=True/ False$: if a node joins a community, it sets $join=True$, otherwise $False$;
\item $head=True/False$: if a node becomes community head, it sets $head=True$, otherwise $False$;
\item $parent=True/False$: if a node becomes parent, it sets $parent=True$, otherwise $False$;
\item $CL(v)$: represents a list of communities where $v$ belongs to at the same time.  Each entry in the list $CL(v)$ is represented as a 4-tuple:  $\{c\_id,c\_size,p\_id, CM\}$, where $c\_id$ and $c\_size$ are defined above, $p\_id$ is parent ID or NULL, and $CM$ is the \textit{community modularity}, respectively.
\item $\mathcal{Z}(v)$: a set of overlapped communities $\{\mathcal{Z}_1,\mathcal{Z}_2,\ldots,\mathcal{Z}_{s}\}$ identified by $v$.
\end{itemize}
\subsection{Problem Definition}
Let $G(V, E)$ be an undirected unweighted underlying network graph, where $V$ is a set of vertices or nodes, and $E$ is a set of edges. 
Our target is to partition $V$ into $\{C_1,C_2,\ldots,C_k\}$ which consists of $k$ number of communities such that $V=C_1\cup C_2\cup\ldots \cup C_k$.

The problem is to identify the overlapped communities $\mathcal{Z}= \{\mathcal{Z}_1,\mathcal{Z}_2,\ldots,\mathcal{Z}_{s}\}$ such that a node
 $ v \in \mathcal{Z}_t$ if $ v \in \displaystyle \cap_{i\in I_t} C_i$ where  $v\in V$, $t \leq s$  and $I_t \subseteq \{1, 2, \ldots, k\}$ with $|I_t| \geq 2$.

\section{ \texttt{DOCD} Algorithm}
\label{sec:algo}
In this section we explain \textsl{Distributed Overlapped Community Detection} (\texttt{DOCD}) algorithm to find the overlapped communities in a network graph $G(V,E)$.

\subsection{Outline of the \texttt{DOCD} algorithm}
\begin{figure}[h!]
  \centering
   \includegraphics[width=0.4\textwidth]{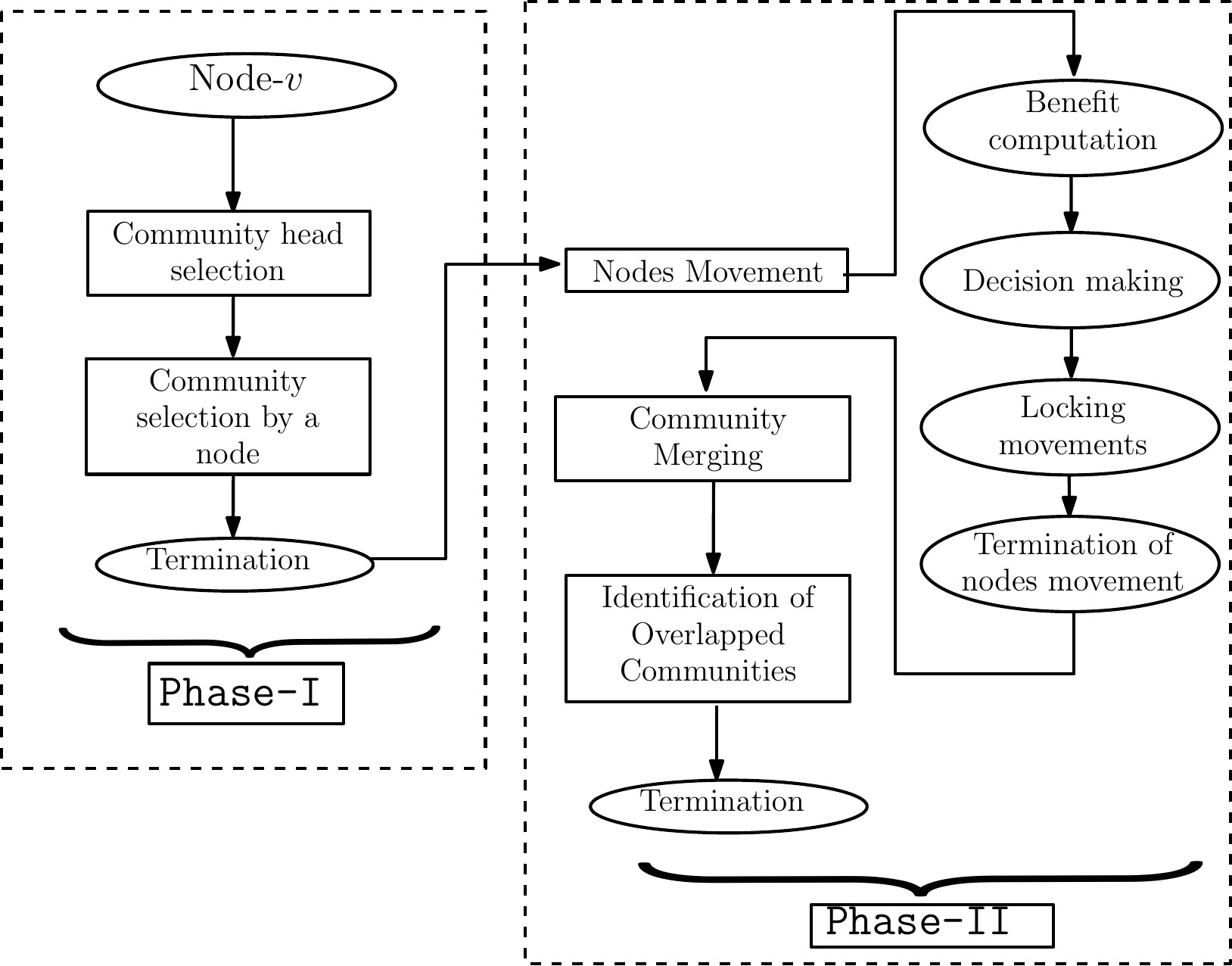}
   \caption{Outline of the \texttt{DOCD} algorithm}
   \label{fig:outline}
\end{figure}

The \texttt{DOCD} algorithm executes in two phases: \texttt{Phase-I} (Community selection and formation) and  \texttt{Phase-II} (Community reorganization). In the first step of \texttt{Phase-I}, community heads are selected, and each community head forms a new community. In the next step, each non-community head selects one or more communities and join them. This community selection process continues until all the nodes join at least one community. \texttt{Phase-I} terminates after all nodes join their respective communities. In  \texttt{Phase-II}, all non-community heads join one or more new communities or leave from its selected communities. Next, the community heads do the same process to leave their community or join a new community. In the case of joining a new community refers to the community merging process. Finally, after the community merging process, each node identifies the overlapped communities. A flow diagram of the \texttt{DOCD} algorithm is illustrated in Fig. \ref{fig:outline}.

A detailed description of the \texttt{DOCD} algorithm is given below.

\subsection{\texttt{ Phase-I:} Community Selection and Formation} 
In this phase, following messages are to be exchanged between the nodes.
\begin{itemize}
\item \texttt{CC\_msg$\langle CC_v \rangle$}: 
\begin{itemize}
\item \texttt{Send:} Node $v$ sends this message to the neighbour after computing the cluster coefficient $CC_v$.
\item \texttt{Receive:} On receiving these messages from all of its neighbour $u$, node $v$ compares its $CC_v$ with $CC_u$.
\end{itemize} 

\item \texttt{Join\_Com$\langle v,c\_id,p\_id\rangle$}:
\begin{itemize}
\item \texttt{Send:} On joining a community a node $v$ sends this message with its community id $c\_id$, and parent id $p\_id$.
\item \texttt{Receive:} On receiving this message, node $v$ updates its $\gamma(v)$.
\end{itemize}

\item  \texttt{Complete$\langle v, NM_v,c\_size \rangle$}:
\begin{itemize}
\item \texttt{Send:} After all neighbours of $v$ select their communities, $v$ sends this massage to its parent. 
\item \texttt{Receive:} On receiving this message, $v$ checks its $\gamma(v)$ to verify that all of its neighbours select their communities or not.
\end{itemize} 
\end{itemize}
The detail description of \texttt{Phase-I} is described below.

\paragraph{\bf Community head selection}
Each node $v$ computes cluster coefficient $CC_v$ using $\Gamma(v)$. It exchange \texttt{CC\_msg$\langle CC_v\rangle $} with its all neighbours. It finds the maximum among all cluster coefficients received from its neighbours (excluding its own) and stores in $CC_{max}$. If its own cluster coefficient, $CC_v > CC_{max}$ then $v$ elects itself as a community head. In case of multiple maximum, id can be used to select the community head. The community head $v$ updates the following information:  

\begin{itemize}
\item The community head inserts a $4$-tuple:  $\{c\_id,c\_size,p\_id, CM\}$ with 
$c\_id=v$, $c\_size=1$, $p\_id=NULL$, and $CM=CC_v$ in its $CL(v)$.
\item It sets $head=True$, $parent=False$ and $join=True$.
\end{itemize}

Finally, $v$ sends a \texttt{Join\_Com$\langle v,c\_id,p\_id \rangle$} message to its neighbours.
Each community head maintains a list of all nodes belongs to its community.

\paragraph{\bf Community selection by a node}
On receiving \texttt{Join\_Com}$\langle v,c\_id,p\_id \rangle$  message, node $u$ first includes the 3-tuple: $(c\_id,-1,-1)$ in   $\gamma(u)$. If it receives multiple \texttt{Join\_Com} messages, then it selects the community which consists of maximum number of its neighbours. $u$ selects multiple communities in case of a tie - having multiple maximum neighbours. In this case, $u$ becomes an \textit{overlapped} node and it will have more than one parents. 

For each selected community, node $u$ inserts an individual entry of a $4$-tuple:  $\{c\_id$, $c\_size,p\_id, CM\}$ with 
$c\_id=v$, $c\_size=1$, $p\_id=v$, and $CM=CC_u$ in its $CL(u)$.
It sets $head=False$, $parent=False$ and $join=True$.
Finally, it sends \texttt{Join\_Com$\langle u,c\_id,p\_id \rangle$} message.
When a node $v$ (already joined a community) receives \texttt{Join\_Com$\langle u,c\_id,p\_id \rangle$} message from node $u$, then node $v$ checks whether it becomes a parent of $u$ or not. Node $v$ checks the $p\_id$ value, if $p\_id==v$ then it updates $parent=True$. 

 \paragraph{\bf Termination of \texttt{Phase-I}}

If a node $v$ is a not a parent (i.e., $parent=False$), then it checks its $\gamma(v)$ to ensure whether all of its neighbours joined their respective communities or not. If joined, it computes the node modularity $NM_v$ by eqn~\ref{eqn:nm}. If $v$ is an overlapped node, it computes $NM_v$ for the individual communities. It sets $c\_size=1$ and sends \texttt{Complete$\langle v, NM_v, c\_size \rangle$} message to its parent then it locally terminates \texttt{Phase-I}.

Now, if a node $v$ is a parent (i.e., $parent=True$), it waits until it receives all the \texttt{Complete} messages from its children. If the parent node $v$ is an an overlapped node and received all the \texttt{Complete} messages from its children then for each community, it does the following tasks.
\begin{itemize}
\item Node $v$ computes $NM_v$ and $NM_{avg}=\frac{NM_v+\sum_{i=1}^l NM_i}{l+1}$, where $NM_i$ is obtained from the  \texttt{Complete} message of neighbour $i$, $l$ is the total number of its children in the same community.

 \item $v$ updates $c\_size= 1+ \sum_{i=1}^{l} c\_size_i$ and sends \texttt{Complete$\langle v, NM_{avg}, c\_size \rangle$} message to its parent.

\item If $v$ is a community head (i.e., $p\_id=NULL$), it computes $NM_v$ and $CM=\frac{NM_v+\sum_{i=1}^{l'} NM_i}{l'+1}$, where $NM_i$ is obtained from received \texttt{Complete} message of neighbour $i$, and $l'$ is the total number of its children. It updates $c\_size=1+\sum_{i=1}^{l'} c\_size_i$.
\end{itemize}
Finally, $v$ locally terminates \texttt{ Phase-I}. 
When each community head receives \texttt{Complete} message from its neighbours, it starts the execution of \texttt{Phase-II} procedure, which is described below.

\subsection{\texttt{Phase-II}: Community Reorganization}
This phase is divided into two sub-phases: \texttt{Nodes movement} and \texttt{Community Merging} which are discussed below.
Here we introduce the following messages those are to be transmitted during the execution of the process.
\begin{itemize}
\item \texttt{Movement$\langle v,CM,c\_size \rangle$}:
	\begin{itemize}
		\item \texttt{Send:} Each community head $v$ sends this message to its community members for a possible movement from one community to another.
		\item \texttt{Receive:} on receiving the message, each member computes the benefits for its current community from which it moves or joins other communities where its neighbours belong to.
	\end{itemize}

\item \texttt{ONM\_msg$\langle ONM_v \rangle$}: 
    \begin{itemize}
        \item \texttt{Send:} Node $v$ sends this message to the neighbour after computing the overlapped node modularity $ONM_v$.
        \item \texttt{Receive:} On receiving these messages from all of its neighbour $u$, node $v$ compares its $ONM_v$ with $ONM_u$.
    \end{itemize} 

\item \texttt{Decision$\langle v,NM_v,benefit,leave\rangle$}:
	\begin{itemize}
		\item \texttt{Send:} When a node  $v$ decides to join or leave a community, it sends this message. A variable $leave=True/False$, is used to know whether it is leaving or not from a community.
		\item \texttt{Receive:} On receiving this message, the community head decides whether its members are allowed to leave or not. 
	\end{itemize}
	
    \item \texttt{Leave\_Accepted $\langle v,CM,c\_size\rangle$}:
	\begin{itemize}
		\item \texttt{Send:} When a community head $u$ decides to allow node $v$ to leave from its community, it sends this message. 
		\item \texttt{Receive:} On receiving this message, a node $v$ leaves a community. 
	\end{itemize}

	\item \texttt{Join\_Req$\langle v,NM_v\rangle$}:
	\begin{itemize}
		\item \texttt{Send:} $v$ sends the message when it decides to join neighbour community. 
		\item \texttt{Receive:} On receiving this message, a node $v$ sends this message to community head. When the community head receives this message, it updates $CM$ and $c\_size$. 
	\end{itemize}
	
\end{itemize}

\subsubsection{\textbf{Nodes Movement}}
\noindent Each community head starts execution of \texttt{Nodes movement} by sending a \texttt{Movement} message to the members of its community. On receiving this message, each member $v$ forwards this message. If $c\_id$ of the message is same with its own community then it updates $CM$ and $c\_size$ in $CL(v)$. If it is not same, then the message came from its neighbour with different communities, then it updates $CM$ and $c\_size$ in $\gamma(v)$. Next, it does the following tasks.
\paragraph{\bf Benefit computation} Node $v$ computes \textit{benefit} to include itself for each of the communities in $\gamma(v)$ and stores the maximum positive benefits with the community id in a list $BL_{nbr}$.
Similarly, it computes \textit{benefits} to exclude itself for each of its communities in $CL(v)$. It selects the maximum positive benefit and stores in a list $BL_{self}$. In case of multiple maximum, it keeps all maximum benefits associate with each of the community ids in $BL_{self}$ or in $BL_{nbr}$ as per the aforementioned explanation.
The detail description of benefit computation is given below:

Let $C_i$ and $C_j$ be two communities with size $l$ and $l'$ respectively. Let $u \in C_i$ and $v \in C_j$ and $u$ and $v$ are neighbours of each other. $u$ computes \textit{benefit} of community modularity for the community  $C_i \setminus \{u\}$ (excluding $u$ from $C_i$) and  $C_j\cup \{u\}$ (including $u$ into $C_j$), respectively.
If $u$ leaves $C_i$, the community modularity of $C_i$ may increase or decrease. Similarly, if $u$ joins $C_j$, that may increase or decrease the community modularity of $C_j$. The community modularity of $C_i$ and $C_j$ are 
$CM_{C_i}$ and $CM_{C_j}$. $u$ computes the \textit{benefit} of $C_i$ after excluding $u$ from $C_i$:\\ $benefit_{(C_i \setminus \{u\})}=CM^{(-u)}_{C_i}-CM_{C_i}~where~ CM^{(-u)}_{C_i}=\frac{(CM_{C_i}\times l)-NM_{u}}{l-1}$.
Similarly, $u$ computes the \textit{benefit} of $C_j$ after including $u$ into $C_j$: $benefit_{(C_j \cup \{u\})}=CM^{(+u)}_{C_j}-CM_{C_j}~ where ~CM^{(+u)}_{C_j}=\frac{(CM_{C_j}\times l')+NM_{u}}{l'+1}$.

\paragraph{\bf Decision making} Movement of $v$ is decided based on following conditions:  
\begin{itemize}
\item if $BL_{self}$ and $BL_{nbr}$ both are empty then $v$ maintains the status quo with current communities. 
\item if $BL_{self}$ is non-empty and $BL_{nbr}$ is empty then $v$  maintains the status quo with current communities.
\item if $BL_{self}$ is empty and $BL_{nbr}$ is non-empty then $v$ joins the neighbour communities along with maintains the status quo with current communities.
\item if $BL_{self}$ and $BL_{nbr}$ both are non-empty then $v$ decides to leave from its current communities and joins neighbour community of positive benefit.
\end{itemize}

\paragraph{\bf Locking movements}
Suppose a node $u$ computes the benefit for its neighbour $v$'s community and decides to join that community. Similarly, node $v$ computes benefits 
for its neighbour $w$'s community and decides to join that community.
If $u$ and $v$ both move to their intended communities, then the benefit computation by node $u$ is incorrect because of the movement of $v$.
To eradicate this problem, we use a locking strategy
by which among the neighbouring nodes, only one node is allowed to move. To execute it, $u$ and $v$ both compute $ONM_u$ and $ONM_v$ by eqn \ref{eqn:onm} and exchange it via the \texttt{ONM\_msg} message. In general, on the comparison, if $ONM_u$ is the minimum \footnote{Minimum value of $ONM_u$ signifies that most of the neighbours of $u$ are belonging to different communities.} among all other competitors (e.g., $v$), then node $u$ is allowed to move. Other competitors (e.g., $v$) lock their movement for this round. If $ONM_u = ONM_v$ for all $v$, then the id of the nodes can be used to break the tie and give priority to one node.
If a node has at least one child having degree one in the community then it also locks itself for the movement. A node $u$ having minimum $ONM_u$, sends a \texttt{Decision$\langle v,NM_v,benefit,leave\rangle$} message to its community head, where $benefit=benefit_{(C_i \setminus \{u\})}$ and  $leave=True$, respectively.

\paragraph{\bf Termination of  nodes movement} 
On receiving \texttt{Decision} message from the members, each community head $v$ makes a list $ML$ of the members who want to leave the community (i.e., whose $leave=True$). $v$ arranges the list $ML$ in decreasing order based on the values of $benefit$. $v$ recomputes the community modularity $CM'$ excluding first member $u$ of the list $ML$. If $CM' \geq CM$ then community head $v$ allows $u$ to leave the  community, where $CM$ is the community modularity including $u$. Then the community head $v$, removes entry of $u$ from $ML$, decreases the $c\_size$ by one, updates $CM$ by $CM'$ and sends \texttt{Leave\_Accepted $\langle u,CM,c\_size\rangle$} message to all of the community members. The community head $v$ repeats the above process unless the list $ML$ is exhausted.
On receiving a \texttt{Leave\_Accepted $\langle u,CM,c\_size\rangle$} message, a member forwards the message to its neighbour within the community and updates $CM$ and $c\_size$.
When $u$ receives this message, it removes the entry of $4$-tuple associated with the community from $CL(u)$. 
When all the children of $u$ receive this message, they select a neighbour with minimum id as a new parent from the same community and update $CM$ and $c\_size$ accordingly.
All the neighbours of $u$ update their $\gamma$.

When a node $u$ (belonging to a community $C_i$) wants to join a new community $C_j$ of a neighbour $w$, it inserts an entry:  $\{c\_id$, $c\_size,p\_id, CM\}$ with 
$c\_id=j$, $c\_size= \textnormal{size of } C_j$, $p\_id=w$, and $CM=CM$ of $C_j$ in its $CL(u)$. Next, $u$ sends \texttt{Join\_Req$\langle u,NM_u\rangle$} message to the neighbour $w$. If $w$ is not the community head then it forwards this message to its community head. On receiving this message the community head updates $CM$ and $c\_size$. 
A non-head node $u$ terminates the \texttt{Node movement} round once it decides the final decision of its movement: staying in the same community or joining in new communities or leaving from the current community and joining new communities. The community head proceeds for the next round \texttt{community merging} once the decision of \texttt{Node movement} in its own community is over. 

\subsubsection{\textbf{Community Merging}}
The community heads are the candidate for executing community merging procedures. This is similar to the aforementioned \texttt{Node movement} procedure. 
In this process, each community head $v$ computes the benefits and stores in $BL_{nbr}$ and $BL_{self}$. Next, if its $BL_{self}$ and $BL_{nbr}$ both are non-empty then $v$ decides to merge with the neighbouring community of the positive benefits. If node $v$ wants to join a new community $C_j$ of community head $u$, then it sends a \texttt{Merge\_Req$\langle v,c\_id, ONM_v \rangle$} message to $u$, where $c\_id=j$ for $C_j$.
If a node $v$ wanted to join $C_j$, but receives a \texttt{Merge\_Req} message from $u$, in that case, lower id node allows other to join by sending a \texttt{Confirm $\langle v,c_v,t,CM \rangle$} message. Meanwhile, if $v$ receives a \texttt{Merge\_Req} message from $w$, then it ignores the message.
On the other case, if node $v$ does not want to join any community and receives a \texttt{Merge\_Req} message from node $w$, then it sends \texttt{Confirm$\langle v,c\_id,c\_size,CM \rangle$} message to node $w$.
When node $v$ joins community of  $u$  after receiving the \texttt{Confirm}$\langle u$, $c\_id$, $c\_size$, $CM\rangle$, it updates $CM$ and $c\_size$ and sends \texttt{Update\_Com} $\langle v$, $c\_id$, $c\_size$, $CM \rangle$ message to all the nodes in the merge community. Each member after receiving this message, updates its $CL(v), \gamma(v),CM,c\_size$, respectively. This process terminates when no community heads move from one community to another. 
\subsubsection{\bf Identification of Overlapped Communities:}
When a node $v$ joins a community, it keeps this community information in its $CL(v)$.
If $CL(v)$ consists of multiple entries, then node $v$ is an overlapped node.
During the whole execution of the \texttt{Phase-II} procedure, each  $v$ updates its $\gamma(v)$. Node $v$ can easily get its neighbours community information from its $\gamma(v)$. Node $v$ checks its $\gamma(v)$ and extracts the neighbour ids of the same communities, and thus, it can identify the overlapped community $\mathcal{Z}(v)=
\{\mathcal{Z}_1,\mathcal{Z}_2,\ldots,\mathcal{Z}_{s}\}$
locally. The outline of \texttt{DOCD} is presented in Algorithm~\ref{algo:docd1}.

\begin{algorithm}[h]
\scriptsize
\KwIn{Node $v$: $\Gamma(v)$}
\KwOut{$\mathcal{Z}(v)$}
	
\tcp { $v$ \textbf{Executes \texttt{Phase-I}}} 
  \begin{itemize}
  	\item Community head selection\;
  	\item Community selection by a node\;
  	\item Termination of \texttt{Phase-I}\;
  \end{itemize}
\tcp{ \textbf{Executes \texttt{Phase-II}}}
	\eIf{$v$ is a non-head member}
	{
	    \If {receives \texttt{Movement} message}
			{
  				node $v$ executes the following task in sequence:
  				\begin{itemize}
  					\item Benefit Computation\;
  					\item Decision Making\;
  					\item Locking Movements\;
  					\item Termination Nodes Movement.
  				\end{itemize}
  		  	}
    }
	{
		\tcp{$v$ is a community head}
			   send \texttt{Movement} message to the members within the community\;
				\If{receives all \texttt{Decision} messages from all community members}
				{
					Executes \texttt{\ Community Merging}\;
				}
    }
			    
	$v$ identifies overlapped communities $\mathcal{Z}(v)$\;
			   
       	\caption{\texttt{DOCD}}
		\label{algo:docd1}
\end{algorithm}
\section{Complexity Analysis}
\label{sec:complexity}
In the first round of \texttt{Phase-I}, the community heads are selected based on cluster coefficients and ids. In the second round, one-hop neighbours of all the community head select their communities. Similarly, two-hop neighbours decide their communities in the third round, and thus the execution of \texttt{Phase-I} proceeds hop by hop in the network.
If the network diameter is $\cal D$, then we require $({\cal D}+1)$ such rounds to terminate the whole process of \texttt{Phase-I}. Hence, the time complexity of the \texttt{Phase-I} is $O(\cal D)$.
In the whole execution of \texttt{Phase-I}, each node $v$ sends one \texttt{CC\_msg}, one \texttt{Join\_Com}, and one \texttt{Complete} message. In this phase per edge requires constant number of messages which is at most six to be traveled during the execution of whole \texttt{Phase-I} procedure. Hence message complexity is $O(m)$.

In \texttt{Phase-II}, each community head initially sends the \texttt{Movement} message to its community neighbours. The neighbour of the community heads forward this message to its community neighbours and so on. Hence, total $\mathcal{D}$ rounds is required for forwarding the \texttt{Movement} message in the worst case.
Each member of a community computes its benefits and makes decisions for its movement. It needs another round of message exchange to know whether it can move or not.
In the same round, it sends \texttt{Decision} message to the community head and a \texttt{Join\_Req} message to its neighbours. Hence, it takes $2\mathcal{D}$ rounds of message exchange. In the next round, each community head sends a \texttt{Leave\_Accepted} message for allowing a member to move. Similarly, this message is forwarded by each of the members that take $\mathcal{D}$ rounds.
Hence, \texttt{Phase-II: Nodes movement} takes $O(\mathcal{D})$ rounds.
In \texttt{Phase-II: Community merging}, each community head computes its benefits. It needs another round to know whether it can merge with the neighbouring community or not. In the next round, it may send or receive a \texttt{Confirm} message and a \texttt{Update\_Com} message. So, \texttt{Phase-II: Community merging} takes  $O(\mathcal{D})$ rounds.
Hence, time complexity of the \texttt{Phase-II} is $O(\mathcal{D})$.
In the whole process of \texttt{Phase-II}, each community head sends  one \texttt{Movement} message. Hence, it becomes total $|C|$ \texttt{Movement} messages.
Next, the community head may generate total $(c\_size-1)$ \texttt{Leave\_Accepted} messages sent to be among the community members.
Since, there are $|C|$ communities, then in the worst case total $|C|$ \texttt{Merge\_Req} messages and $|C|$ \texttt{Confirm} messages may generate during the whole process.
Finally, total $c\_size$ number of \texttt{Update\_Com} messages may exchange for updating the community.
Similarly, each node in the network, who decides for movement, exchanges a \texttt{ONM\_msg} and a \texttt{Decision} message. Hence, total $2n$ messages are to be generated. Finally, a member can join $|C|-1$ communities, thus $n$ nodes may generate at most $n\cdot (|C|-1)$ \texttt{Join\_Req} messages.
Hence, total message complexity is $O(|C|+c\_size+2n+n\cdot (|C|-1))= O(n\cdot |C|)\approx O(n^2)$, where  maximum possible value of $|C|$ is $n$. In worst case each message can pass through every edge in the network, hence message complexity of \texttt{DOCD} algorithm is $O(n^2m)$ and time complexity is $O(\mathcal{D})$.

\section{Performance Evaluation}
\label{sec:simulation} 

Extensive simulation studies have been done to evaluate the performance of our proposed \texttt{DOCD} algorithm.
\begin{figure}[h]
  \centering
   \includegraphics[width=\linewidth]{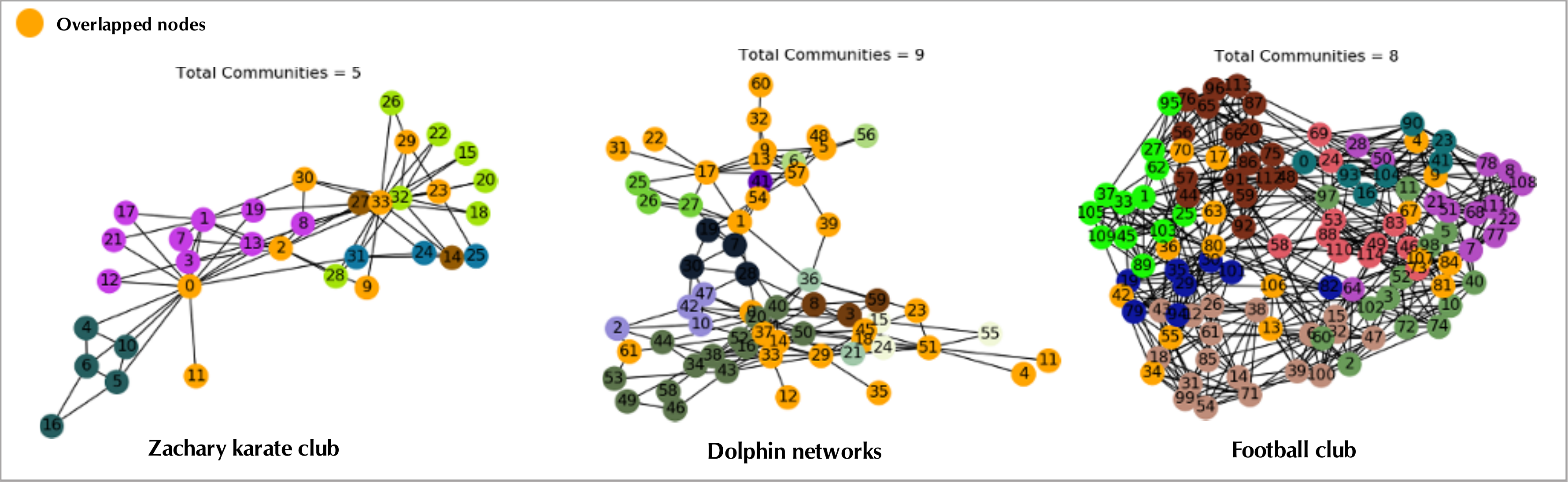}
   \caption{Run time snapshot of real world networks (benchmark datasets)}
   \label{fig:snapshot}

\end{figure}
 In our simulation study, we use real world networks (benchmark datasets) like, \texttt{Dolphin}, \texttt{Zachary karate club}, \texttt{Football club}, and \texttt{Facebook} networks, respectively.
The algorithm is implemented in Intel Xeon 2.6GHz, 16 cores, 64 GB RAM, machine using  Python-3. 
A run-time snapshot of the \texttt{DOCD} algorithm is shown in Fig. \ref{fig:snapshot} for \texttt{Dolphin}, \texttt{Zachary karate club} and \texttt{Football club} networks. 

\begin{figure}[ht]
\centering
\includegraphics[width=0.4\textwidth]{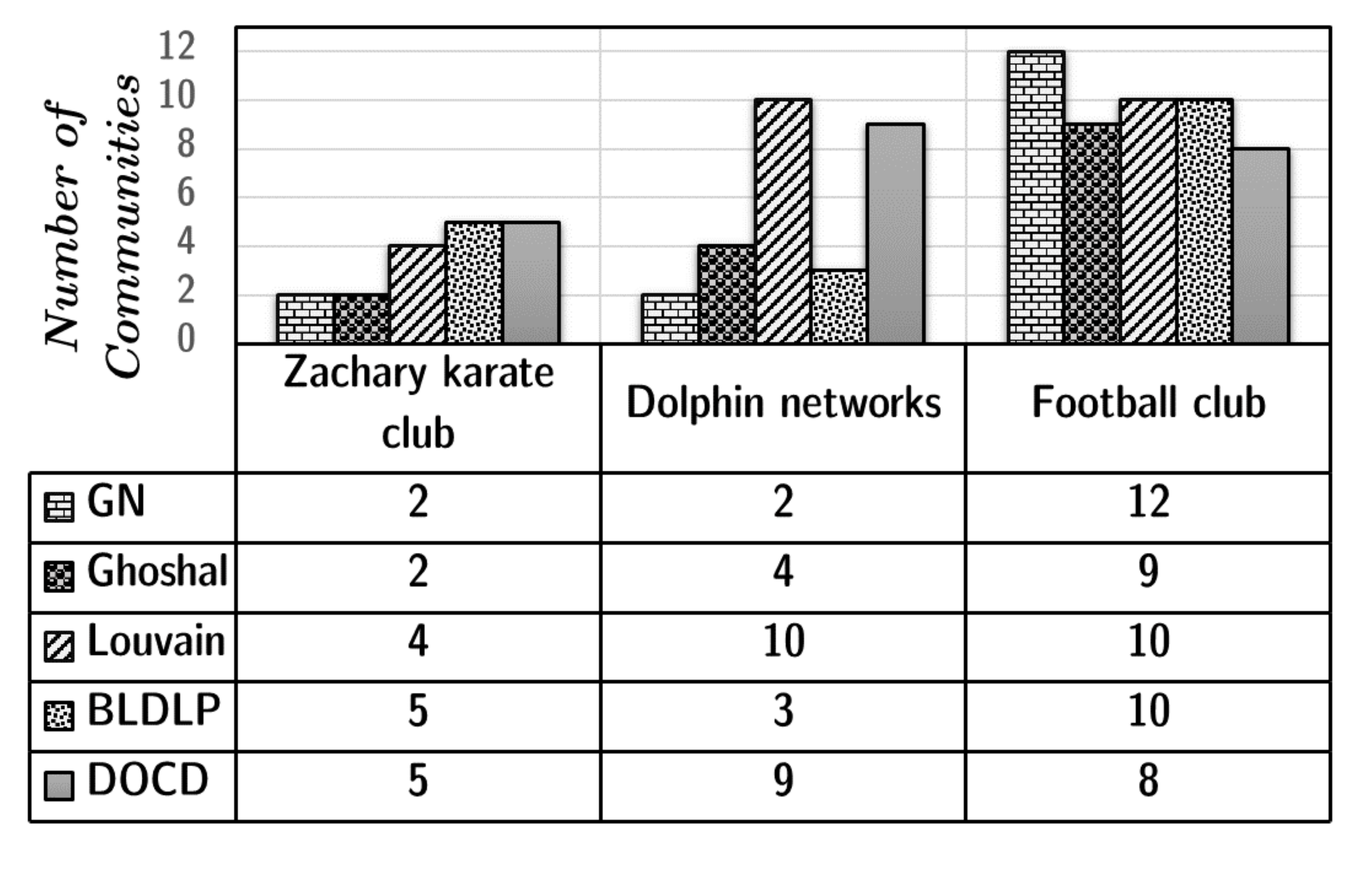}
		\caption{Comparison with number of communities on different networks}
   \label{fig:no_communities}
   \end{figure}

The different and same colored nodes in  Fig. \ref{fig:snapshot} show different and same community, respectively. All the orange-colored nodes refer to the overlapped nodes.

\begin{figure}[!h]
\centering
\includegraphics[width=0.4\textwidth]{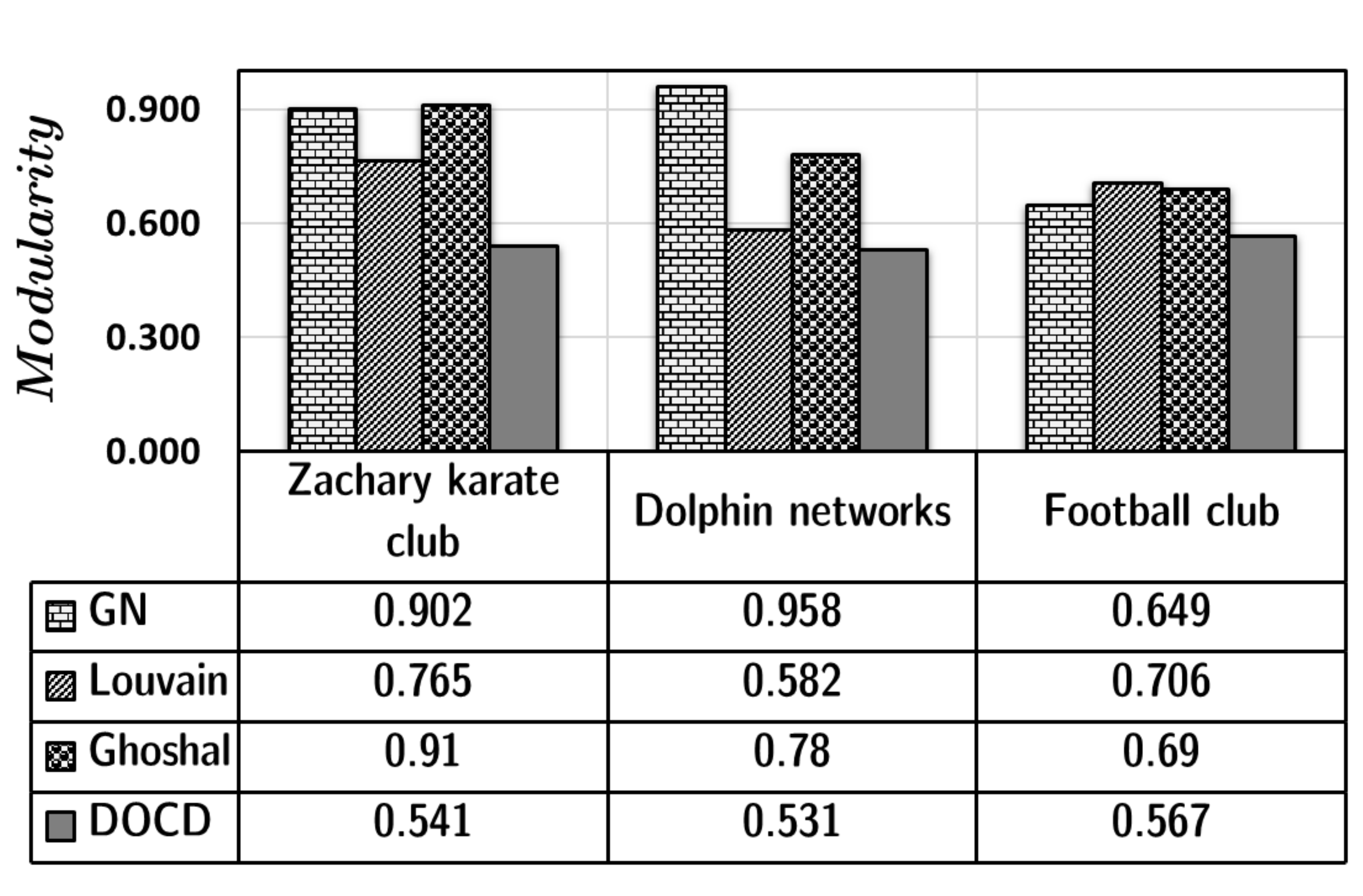}
    \caption{Comparison with community modularity on different networks}
    \label{fig:modularity}
\end{figure}

Fig. \ref{fig:no_communities} shows number of identified communities on  \texttt{Dolphin}, \texttt{Zachary karate club} and \texttt{Football club} networks, respectively. We have compared our results with GN \cite{Newman-2004}, Louvain \cite{Blondel-2008}, Ghoshal \cite{Ghoshal-2017}, and BLDLP  \cite{Jokar-2019} methods. 
 
We observe that \texttt{DOCD} identifies almost the same number of communities as a result obtained by Louvain \cite{Blondel-2008} for all three benchmark datasets, whereas for \texttt{Football club} network, the result of \texttt{DOCD} is almost similar with the outcome of \cite{Ghoshal-2017}.
Hence, \texttt{DOCD}, the algorithm performs almost comparable results compared to the centralized algorithms in terms of identifying number communities with existing works.

\begin{figure}[!h]
\centering
\includegraphics[width=0.4\textwidth]{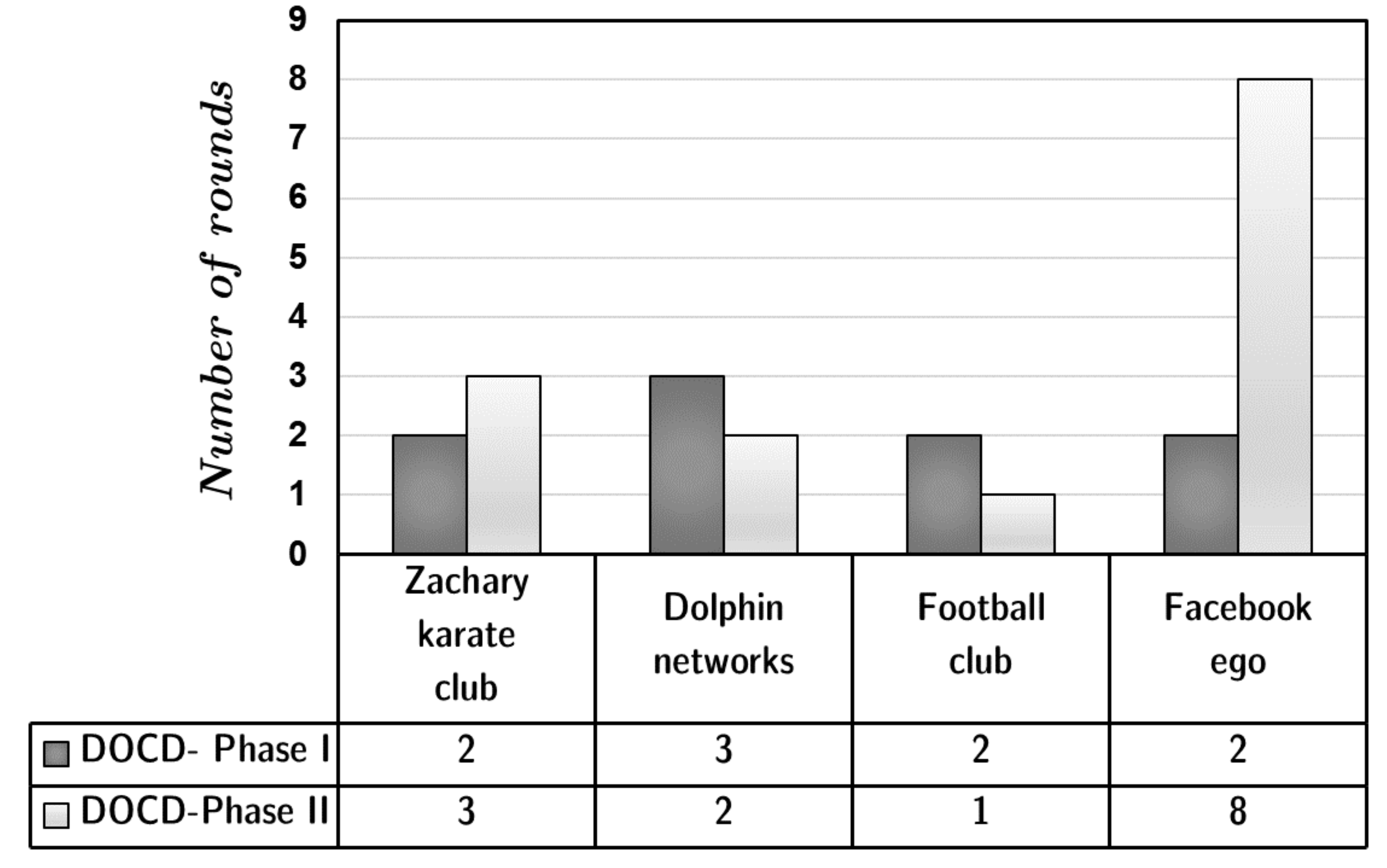}
   		\caption{Number of rounds needed to find overlapped communities for different networks}
   		\label{fig:rounds}
\end{figure}

\begin{figure}[!h]
\centering
\includegraphics[width=0.4\textwidth]{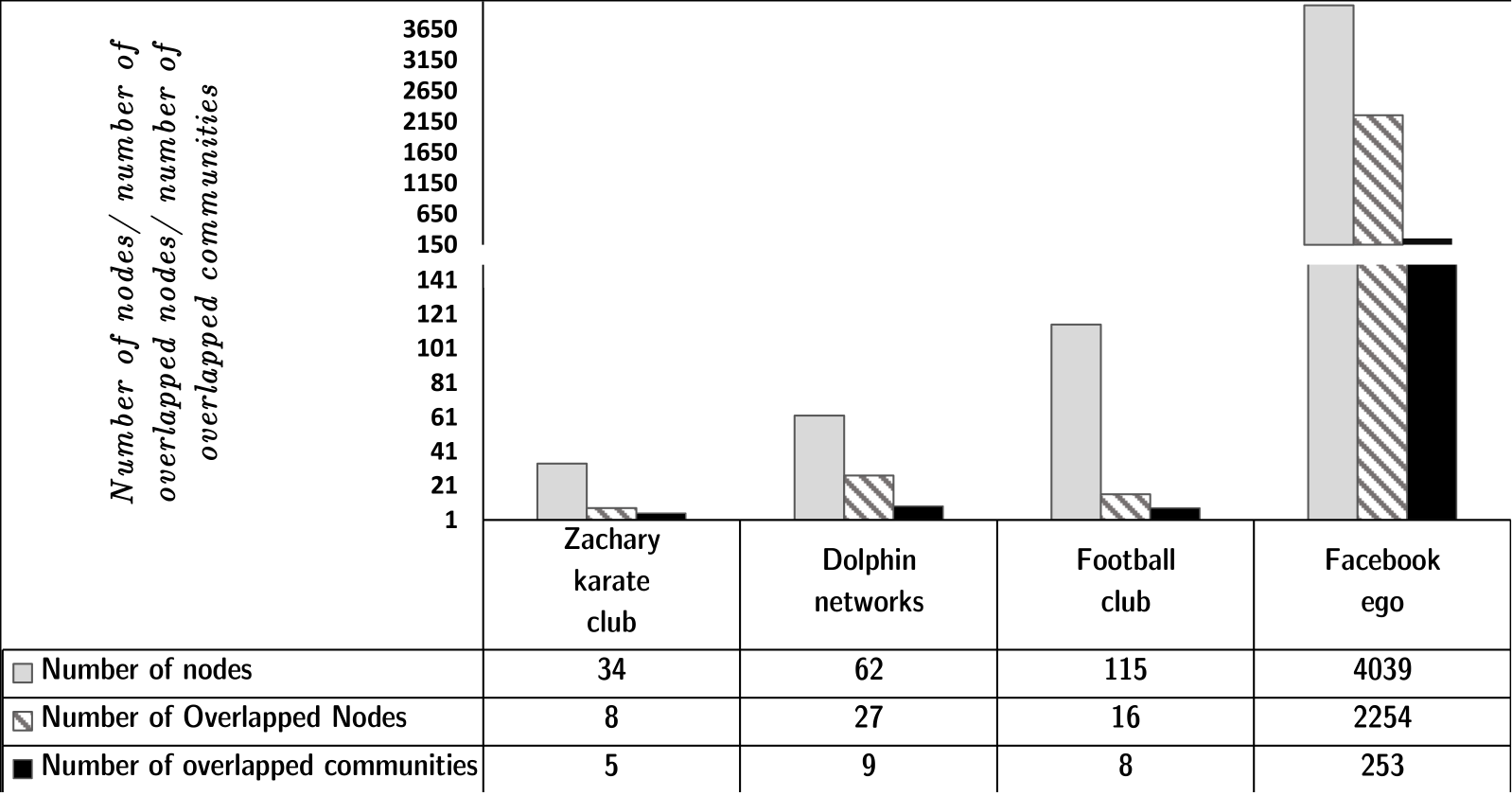}
   		\caption{Number of nodes, number of overlapped nodes and number of overlapped communities in different networks}
  		\label{fig:overlapped_nodes}
\end{figure}
  
Fig. \ref{fig:modularity} shows  community modularity values of different algorithms for  \texttt{Dolphin}, \texttt{Zachary karate club} and \texttt{Football club} networks.
Compare to others, the community modularity value of \texttt{DOCD} algorithms is less. The reason for the same is that \texttt{DOCD} finds overlapped communities where community modularity is computed using overlapped nodes, whereas \cite{Newman-2004}, \cite{Blondel-2008}, and  \cite{Ghoshal-2017} compute disjoint communities.

Fig. \ref{fig:rounds} shows the number of rounds required  to find overlapped communities in \texttt{Zachary karate club}, \texttt{Dolphin} and \texttt{Football club} networks, respectively using \texttt{DOCD} algorithm. We show the results for each phase individually.       
Fig. \ref{fig:overlapped_nodes} shows the variation of the actual number of nodes vs. number of overlapped nodes and overlapped communities identified from the  \texttt{Zachary karate club}, \texttt{Dolphin} and \texttt{Football club} networks, respectively.
\begin{table}[h]
\scriptsize
\centering
\begin{tabular}{|l|p{0.7cm}|p{0.6cm}|p{0.7cm}|p{0.6cm}|p{1cm}|p{0.6cm}|}
\hline
\multirow{2}{*}{Data set} & \multicolumn{2}{p{1.7cm}|} {No. of communities}  & \multicolumn{2}{p{1.7cm}|}{Community modularity} & \multicolumn{2}{p{1.7cm}|}{No. of overlapped nodes}  \\\cline{2-7}                                                            
  & Meena \cite{Meena-2015} & \texttt{DOCD} & Meena \cite{Meena-2015} & \texttt{DOCD} & Meena \cite{Meena-2015} & \texttt{DOCD} \\ \hline
Karate  & 4  & 5   & 0.4198  & 0.541 & 5  & 8   \\ \hline
Dolphin  & 5& 9 & 0.5285 & 0.531 & Unknown  & 27  \\ \hline
Football & 7 & 8  & 0.5851  & 0.567& Unknown  & 16  \\ \hline
\multicolumn{1}{l}{}   
\end{tabular}
\caption{Comparison with \cite{Meena-2015} with respect to the number of communities, community modularity and number of overlapped nodes.}
\label{tab:comp}
\end{table}

Table~\ref{tab:comp} shows the comparison study with the method  \cite{Meena-2015}, for the number of identified communities, community modularity, and the number of overlapped nodes, respectively. It is interesting to observe that the number of identified communities are always greater than the result reported in \cite{Meena-2015}. The comparison of the number of overlapped nodes is insignificant because the number of overlapped nodes directly affects the community modularity value. Thus, our \texttt{DOCD} algorithm results better community modularity as compared to \cite{Meena-2015}
 \begin{table}[!h]
\centering
\scriptsize
\begin{tabular}{|p{1.2cm}||c|p{0.3cm}|c|p{0.3cm}|c|p{0.3cm}|p{1cm}|p{1cm}|}
\hline
Method & \multicolumn{2}{c|}{Zachary} &  \multicolumn{2}{c|}{Dolphin} & \multicolumn{2}{c|}{Football} & Algorithm & Community\\
 & \multicolumn{2}{c|}{Karate club} &  \multicolumn{2}{c|}{networks} & \multicolumn{2}{c|}{club} & &  \\
\cline{2-9}
& $|C|$ & $CM$ & $|C|$ & $CM$ & $|C|$ & $CM$ & \multicolumn{2}{l|}{} \\ 
\cline{1-9}
GN \cite{Newman-2004} & 2 & 0.90 & 2 & 0.95 & 12 & 0.64 & centralized & Disjoint\\
Louvain\cite{Blondel-2008} & 4 & 0.91 & 10 & 0.78 & 10 & 0.69  & Centralized & Disjoint \\
Ghoshal \cite{Ghoshal-2017} & 2 & 0.76 & 4 & 0.58 & 9 & 0.70 & Centralized & Disjoint\\
Meena \cite{Meena-2015} & 4 & 0.41 & 5 & 0.52 & 7 & 0.58 & Centralized & Overlapped\\
DOCD & 5 & 0.54 & 9 & 0.53 & 8 & 0.56 & Distributed & Overlapped \\ \cline{1-9}
\end{tabular}
\caption{showing comparison summary with the number of communities $|C|$ and community modularity $CM$}
\label{tab:summary}
\end{table}

The \texttt{DOCD} algorithm identifies overlapped communities in a large-scale networks and it produces comparative results with the existing centralized algorithms \cite{Newman-2004,Blondel-2008,Ghoshal-2017}, which are shown in Table~\ref{tab:summary}. Moreover, \texttt{DOCD} algorithm is capable of identifying the overlapped communities within a small number of rounds of communications, as shown in Fig. \ref{fig:rounds}. 

\section{Conclusion}
\label{sec:conclusion}
This paper has proposed a distributed algorithm \texttt{DOCD}, to solve the overlapped community detection problem in large-scale networks. The total number of nodes in the networks is not an input of the algorithm. It is scalable and robust with respect to the number of nodes in the networks.
The time and message complexities of the algorithm are $O(n^2m)$ and $O(\mathcal{D})$, respectively, where $m$, $n$, and $\mathcal{D}$ are the number of nodes, edges, and diameter of the network graph. To show the performance of the algorithm, we have done an extensive simulation study with benchmark data sets. We showed that our distributed algorithm keeps the asymptotically same numbers of communities and community modularity with existing centralized algorithms. 
The proposed solution can also work for a dynamic network by recomputing community modularity for a neighborhood change, but complexity will be high. So, as future work, one can design an efficient distributed algorithm for dynamic networks.
\section*{Acknowledgment}
Dibakar Saha would like to acknowledge the Science
and Engineering Research Board (SERB), Government of
India, for financial support under the NPDF scheme (File
Number: PDF/2018/000633).

\balance
\bibliographystyle{IEEETran}
\bibliography{ref}
\end{document}